\documentclass[12pt]{article}
\usepackage{epsf}
\catcode `@=11 \@addtoreset{equation}{section} \catcode `@=12
\begin{document}
\setlength{\baselineskip}{7mm}
\def\be{\begin{equation}}
\def\ee{\end{equation}}
\def\ba{\begin{array}{l}}
\def\ea{\end{array}}
\def\bea{\begin{eqnarray}}
\def\eea{\end{eqnarray}}
\def\eq#1{(\ref{#1})}
\def\fig#1{Fig \ref{#1}}
\def\wgnc{\bar{\wedge}}
\def\del{\partial}
\def\der{\overline \del}
\def\wg{\wedge}
\def\bull{$\bullet$}
\def\gap{\vspace{10ex}}
\def\tgap{\vspace{3ex}}
\def\sgap{\vspace{5ex}}
\def\lgap{\vspace{20ex}}
\def\half{\frac{1}{2}}
\def\pto{\vfill\eject}
\def\gst{g_{\rm st}}
\def\tC{{\widetilde C}}
\def\z{{\bar z}}
\def\o{{\cal O}}
\def\J{{\cal J}}
\def\S{{\cal S}}
\def\X{{\cal X}}
\def\N{{\cal N}}
\def\A{{\cal A}}
\def\H{{\cal H}}
\def\D{{\tilde D}}
\def\d{{\cal D}}
\def\re#1{{\bf #1}}
\def\nn{\nonumber}
\def\nl{\hfill\break}
\def\ni{\noindent}
\def\bibi{\bibitem}
\def\c#1{{\hat{#1}}}
\def\eps{{\epsilon}}
\pretolerance=1000000
\begin{flushright}
KEK/TH/722\\
October 2000\\
\end{flushright}
\begin{center}
\vspace{2 ex}
{\large\bf Wilson loops in strongly coupled noncommutative gauge theories}
\\
\vspace{3 ex}
Avinash Dhar $^{1*}$ and Yoshihisa Kitazawa $^{\dagger}$ \\
~\\
{\sl Laboratory for Particle and Nuclear Physics,}\\
{\sl High Energy Accelerator Research Organization (KEK),}\\
{\sl Tsukuba, Ibaraki 305-0801, JAPAN.} \\

\vspace{10 ex}
\pretolerance=1000000
\bf ABSTRACT\\
\end{center}
\vspace{1 ex}

We discuss Wilson loop averages in $4$-dimensional non-commutative
superYang-Mills theory using the dual supergravity description.
We postulate that the Wilson loops are
located at the mimimum length scale $R$ in the fifth radial coordinate.
We find that
they exhibit a crossover from Coulomb type of behaviour for large loops, for
which non-commutativity is unimportant, to area law for small loops, for which
non-commutativity effects are large.  The string tension, which can be read off
from the area law, is controlled by the non-commutativity scale.  The crossover
itself, however, appears to involve loops of size of order $R$
which is much larger than the non-commutativity scale.  The existence of
the area law in non-commutative super Yang-Mills theory
which persists up to a large crossover length scale
provides further
evidence for connection to an underlying string theory.

\vfill
\hrule
\vspace{0.5 ex}
\leftline{$^1$ On leave from Dept of Theoretical Phys, Tata Institute, Mumbai
400005, INDIA.}
\leftline{$^*$ adhar@post.kek.jp}
\leftline{$^\dagger$ kitazawa@post.kek.jp}

\clearpage

\vspace{8ex}

\section{\large Introduction}

Non-commutative Yang-Mills (NCYM) theories arise as certain limits of closed
string theories, in the presence of D-branes and background B-field.  Recently
these theories have attracted a lot of attention
\cite{AIIKKT,C1,I,IIKK1,BM,AMNS,MRS,AD,GW,F,RU,DR,GHI,DW}, partly because
non-commutative geometry naturally arises in string theory
\cite{CDS,DH,AAJ,CK,C2,CH,VS,SW1} and partly because string theory simplifies
considerably in the limit of large non-commutativity
\cite{HKLM,DMR,W,GMS,S,KRS}
allowing a detailed study of a number of important issues.

One question of intrinsic interest in any gauge theory is the issue of the
construction of local gauge-invariant operators since these form a
complete set of observables of the theory.  This question has been
addressed in several recent works.  In \cite{IIKK1} a set of
gauge-invariant operators in NCYM theories was constructed, and these were
further studied from different perspectives in
\cite{AD,RU,DR,GHI,DW}. Roughly speaking, these gauge-invariant
observables can be written as Fourier transforms of open Wilson lines. In
the operator formalism they are given by the following expression
\bea
W_k[C]=Tr\bigg( Pexp\{i\int_C d\sigma \ \del_\sigma
y^\mu(\sigma)A_\mu(\hat x+y(\sigma))\} e^{ik.\hat x} \bigg),
\label{oneone}
\eea
where
\bea
[{\hat x}^\mu, {\hat x}^\nu]=i\theta^{\mu\nu},
\label{onetwo}
\eea
and the trace in (\ref{oneone}) is over both the gauge group as well as
the operator Hilbert space. These open Wilson lines are gauge-invariant in
NCYM theories, unlike in ordinary YM theories, provided the straight line
joining the end points of the path $C$, given by $y^\mu(\sigma)$, is fixed
to be the vector
\bea
l^\mu=k_\mu\theta^{\mu\nu}.
\label{onethree}
\eea
The path $C$ is otherwise completely arbitrary. When $k$ vanishes, the two
ends of the path $C$ must meet and we have a closed Wilson loop.

Recently, multi-point correlation functions of the Wilson line operators
have been computed in NCYM theory by Gross, Hashimoto and Itzhaki
\cite{GHI}. These authors have found a universal exponential suppression of
the normalized correlators at high momenta, reminiscent of the behaviour
of form factors in high energy fixed angle scattering of string
theory. There are other hints of relation between NCYM theory and string
theory. It has, in fact, been suggested that $4$-dimensional ${\cal N}=4$
NCSYM is dual to superstring theory in appropriate backgrounds with the
non-commutativity scale as its string scale \cite{MR,HI,IIKK1,IIKK2,RU}.

In this paper we provide further evidence of this string connection of NCYM
theory. We study the behaviour of closed Wilson loops in NCSYM using the
dual supergravity description \cite{MR,HI} \footnote{Other examples that
discuss non-commutative AdS/CFT correspondence are \cite{NS,SW2,DMWY,
DRT,DG}}. In our investigation,
we postulate that the Wilson loops are located at the
length scale $R$ in the fifth radial coordinate where the warp
factor peaks.
It is certainly very diferent from the standard procedure of
ordinary $AdS$/CFT correspondence.
In that case the Wilson loops are
located at the infinity of the fifth coordinate where the warp
factor diverges.

Let us recall why such a choice is reasonable
in ordinary field theory.
The fifth coordinate has been often interpreted as the
renormalization scale.  In such an interpretation,
we need to rescale the warp factor by $\alpha^2$ if we change
the length scale by $\alpha$.  Such a logic leads to
the conclusion that the local theory
can only be located at where the warp factor diverges.
In NCYM case, the warp factor in noncommutative dimensions
has a maximum at the length scale $R$ in the fifth radial coordinate.
It seems that all we can do to approach the bare theory
is put the Wilson loops at the minimum length scale $R$.
The warp factor can also be interpreted as the effective string tension.
It is reassuring to find that the effective
string tension is equal to noncommutativity scale in our choice.

The organization of this paper is as follows.
We investigate the behaviour of closed Wilson loops in Sec.2.
We find that while sufficiently large Wilson
loops show a Coulomb law behaviour, as confirmed earlier in \cite{MR,AOJ},
there is a crossover to an area law behaviour for small loops. The
crossover is controlled by the scale that enters the supergravity
solution, which is much larger than the non-commutativity scale. However,
the string tension, which can be read off from the area law, is controlled
by the noncommutativity scale. We end with a summary and discussion of our
results in Sec.3.

\section{\large NCSYM Wilson loops in dual supergravity}

In this section we will first summarize the supergravity solution which is
dual to NCSYM theory \cite{MR,HI}. We will then discuss calculation of
Wilson loop averages using this solution.

\subsection{Background solution }

The space-time directions of the NCSYM theory are labeled by $x_0, x_1,
x_2$ and $x_3$. The only non-zero B field component is in the space
directions $x_2, x_3$, resulting in non-commutativity only in these
directions. We give below only the string metric and B field, since the
other fields will not be relevant for the calculations that follow.
\bea
ds^2 &=& \alpha' \sqrt\lambda \bigg[
u^2(- {dx_0}^2 + {dx_1}^2) + u^2 h ({dx_2}^2 + {dx_3}^2) + {du^2 \over
u^2} + d {\Omega_5}^2 \bigg], \label{twoone} \\
h^{-1} &=& 1+u^4R^4, \ \ \ \lambda=4 \pi g N, \ \ \ R^4=\lambda \theta^2,
\label{twotwo} \\
B_{23} &=& \alpha' \sqrt\lambda R^2u^4h.
\label{twothree}
\eea
Here $\theta=\theta^{23}$, the latter being as defined in (\ref{onetwo}).
The metric reduces to the standard AdS solution for $uR<<1$. The scale $R$
that enters here is different from the non-commutativity scale defined by
$\theta$ and, in fact, is much larger than the latter because the 't Hooft
coupling $\lambda$ must be large for a consistent supergravity
solution. The fact that the effects of non-commutativity persist at scales
much larger than that defined by $\theta$ has been interpreted in the
literature \cite{MR,HI} as renormalization of the non-commutativity scale due
to strong coupling effects. However, as we shall see below, the scale $R$
is perhaps more correctly interpreted as the scale which controls the
crossover behaviour of Wilson loops from Coulomb to area law.

\subsection{Wilson loop averages}

In the partially non-commuting case described by the supergravity solution
(\ref{twoone}-\ref{twothree}) there can be three distinct kinds of Wilson
loops. Those that lie entirely in the commuting $x_0-x_1$ plane, those
that lie entirely in the non-commuting $x_2-x_3$ plane, and those that lie
in mixed commuting and non-commuting directions. Wilson loops of the first
kind always lead to Coulomb law, just like the AdS/CFT case involving
ordinary $4$-dimensional SYM theory. Wilson loops of the second and third
kind show a more interesting behaviour and it is these type of loops that
we will now consider.

In the case of ordinary $4$-dimensional SYM theory in the strong coupling
limit, Wilson loop averages are postulated to be given by minimizing the
relevant Nambu-Goto area action \cite{RY,M}. In the present case of NCSYM
theory, there is also a non-zero B field and, for the generic Wilson loop,
the B field term also contributes in the string action. Thus, the relevant
action to consider here is the area action together with the B field term:
\bea
S={1 \over 2\pi \alpha'} \int d\tau d\sigma \bigg[\sqrt{-{\rm det} g} +
B_{\mu\nu} \del_\tau x^\mu \del_\sigma x^\nu \bigg],
\label{twofour}
\eea
where
\bea
g_{\alpha\beta}=G_{\mu\nu} \del_\alpha x^\mu  \del_\beta x^\nu, \ \
\alpha, \beta = \tau, \sigma
\label{twofive}
\eea
is the induced metric on the world-sheet. As in \cite{AOJ}, we will consider
the string configuration
\bea
x_0=\tau, \ x_3=\sigma, \ x_2=v\tau, \ u=u(\sigma).
\label{twosix}
\eea
Thus, the string is moving in the $x_2$ direction with velocity $v$. As a
result, on any $4$-dimensional fixed $u$ slice, the contour of the loop is
not along coordinate directions. The loops are, however, rectangular with
either one time-like and one space-like direction or both directions
space-like, depending on what we want. Thus, from the above string
configuration we can either get mixed loops with one commuting and one
non-commuting direction, or the space-like loops lying entirely in the
non-commuting $x_2-x_3$ plane.

For the string configuration (\ref{twosix}) the action (\ref{twofour}) is
\bea
S={T \sqrt\lambda \over 2 \pi} \int d\sigma \bigg[
\sqrt{(1-v^2h)({u'}^2+u^4h)} + vR^2u^4h \bigg],
\label{twoseven}
\eea
where $u'=\del_\sigma u$. At an extremum of this action the quantity
$$u^4h \bigg({\sqrt{1-v^2h} \over \sqrt{{u'}^2+u^4h}} + vR^2 \bigg)$$ is
independent of $\sigma$. This gives the solution
\bea
u'={bu^2 \sqrt{u^4-u_0^4} \over u_0^2(b+cu^4/u_0^4)},
\label{twoeight}
\eea
where the constants $b$ and $c$ are given by
\bea
b &=& \sqrt{1-v^2h_0} + vR^2u_0^2 \sqrt{h_0},
\label{twoten} \\
c &=& R^4u_0^4 \sqrt{1-v^2h_0} - vR^2u_0^2 \sqrt{h_0}.
\label{twoeleven}
\eea
The constant $u_0$ is the turning point in $u$ where $u'$ vanishes and
$h_0$ is the value of $h$ at $u=u_0$. From the solution (\ref{twoeight})
we obtain
\bea
x_3(u)={1 \over bu_0} \int_1^{u/u_0} dy \ {b+cy^4 \over y^2 \sqrt{y^4-1}}.
\label{twotwelve}
\eea
For $v=0$, $b=1$ and $c=R^4u_0^4$. This is the case considered in
\cite{MR}. In this case $x_3(u)$ grows linearly with $u$ at large
values. However, for the special value $v=R^2u_0^2$, $c$ vanishes \cite{AOJ}
and then $x_3$ remains finite as $u$ becomes large.

We will work here with generic values of $v$.  This is because from the
viewpoint of supergravity calculations, $x_3(u)$ has no direct physical
significance.  The physical quantity is $x_3(u)/uR$, which measures proper
distances along this direction at large $u$, and this remains finite as $u
\rightarrow \infty$.  In fact, at large $u$ the metric in (\ref{twoone}) has
the
scaling isometry
\bea
x_{0,1} \rightarrow \beta^{-1} x_{0,1}, \ x_{2,3} \rightarrow \beta
x_{2,3}, \ u \rightarrow \beta u.
\label{twothirteen}
\eea
This also shows that $x_3(u)$ has no direct physical relevance at large $u$.
However, the physical quantity $x_3(u)/uR$ is invariant under the above scaling
isometry.

Another way of saying the same thing is the following.  Let us take the
boundary
to be the $4$-dimensinal slice of the metric (\ref{twoone}) at $u=\Lambda$.
The
metric on this $4$-dimensional slice is given by
\bea
ds_4^2=\alpha' \sqrt\lambda \bigg[ \Lambda^2(- {dx_0}^2 + {dx_1}^2) +
{\Lambda^2 \over (1+R^4\Lambda^4)} ({dx_2}^2 + {dx_3}^2) \bigg].
\label{twofourteen}
\eea
For finite coordinate distances in the non-commuting directions, proper
distances in these directions become very small, even smaller than string
scale,
at large $\Lambda$.  Since we are neglecting stringy corrections in the present
semiclassical analysis, we cannot trust results obtained in this case.
However,
for coordinate distances growing linearly with $\Lambda$, proper distances
remain finite at large $\Lambda$ and so the present approximation is reliable.
This suggests that we should consider Wilson loops which have a large
horizontal
length.  An alternative possibility is that the ``boundary'' might be
located at
a finite value of $\Lambda$.  As we shall see in the following, there are
actually strong reasons for considering the latter possibility.

\subsection{Rectangular Wilson loops with a time-like direction}

Let us now consider the case of rectangular Wilson loops with a time-like
direction. At any constant $u$ slice, the length of the loop in the
time-like direction equals $\sqrt{1-v^2h}$. For this to be real and
non-zero, we require $(1-v^2h)>0$. This is satisfied for all values of $u$
if we take $|v|<1$. In this subsection we will restrict ourselves to this
range of values of $v$ \footnote{The value $v=1$ is singular in the sense
that the action vanishes on the classical solution.}.

Now, let $L$ be the length between the two ends of the string in the $x_3$
direction on the $4$-dimensinal slice at $u=\Lambda$. From the solution
(\ref{twoeight}) we get,
\bea
{L \over 2}={1 \over bu_0} \int_1^{\Lambda/u_0} dy \ {b+cy^4 \over y^2
\sqrt{y^4-1}},
\label{twofifteen}
\eea
and the action evaluated on the solution is
\bea
S={T \sqrt\lambda \over \pi }(b+c)u_0 \sqrt{h_0}
\int_1^{\Lambda/u_0}dy{y^2 \over \sqrt{y^4-1}}.
\label{twosixteen}
\eea

We should mention that for $|v|<1$ while $b$ is always positive, $c$ is
positive
only for $v<R^2u_0^2$ and changes sign for $v>R^2u_0^2$.  If $Ru_0<1$, this is
within the range of values of $v$ being considered here.  In this case the
denominator in (\ref{twoeight}) vanishes at $u=|b/c|^{1 \over 4}$.  Beyond this
point the string bends back and crosses itself at some larger value of $u$.  It
would be interesting to understand the physics of such self-intersecting
surfaces.  Here we will avoid such configurations by restricting ourselves to
$v<R^2u_0^2$ for $Ru_0<1$.  If $Ru_0>1$, these configurations do not occur
because of the restriction to $|v|<1$.

Let us now analyse the expressions (\ref{twofifteen}) and (\ref{twosixteen}).
Two different cases arise which we will now discuss in turn.

\ni {\bf case(i):$u_0 << \Lambda$}

In this case (\ref{twofifteen}) and (\ref{twosixteen}) give
\bea
{L \over 2} &=& {1 \over bu_0} \bigg[c{\Lambda \over u_0}+(b-c)
{\pi \sqrt{2 \pi}
\over \Gamma({1 \over 4})^2}+ \cdots \bigg],
\label{twoseventeen} \\
S &=& {T \sqrt\lambda \over \pi }(b+c)u_0 \sqrt{h_0} \bigg[{\Lambda
\over u_0}-{\pi \sqrt{2 \pi} \over \Gamma({1 \over 4})^2}+ \cdots \bigg].
\label{twoeighteen}
\eea
The dots in these expressions refer to terms that are suppressed by powers of
$\Lambda$.  Now, if the location $u=\Lambda$ of the boundary is held fixed,
then
for an appropriately small value of $Ru_0$ ($<<1$) the second term in
(\ref{twoseventeen}) can be made much larger than the first term.  To be
concrete, let us take $v=0$.  Then, $b=1$ and $c=R^4u_0^4$.  The expression for
$L$ then becomes
\bea
{L \over 2}={1 \over u_0} \bigg[R^4u_0^3 \Lambda +(1-R^4u_0^4)
{\pi \sqrt{2 \pi} \over \Gamma({1 \over 4})^2}+ \cdots \bigg].
\label{twoeighteena}
\eea
We see that the second term dominates for $Ru_0 << (R\Lambda)^{-1/3}$.  In this
case, after subtracting the W-boson mass contibution, we recover the Coulomb
interaction familiar from AdS/CFT duality of ordinary SYM theory:
\bea
S \approx -{4 \pi^2 \sqrt{\lambda} \over \Gamma({1 \over 4})^4} \ {T \over L}
\label{twonineteen}
\eea
This is as expected since loops with large values of $L$ should not see the
effects of non-commutativity.

An interesting aspect of the expression (\ref{twoeighteena}) for $L$ is that it
attains a minimum value as a function of $u_0$.  This happens at $Ru_0 \sim
(R\Lambda)^{-1/3}$ for which $L \sim R(R\Lambda)^{1/3}$.  For values of $L$
above this minimum, there are two different values of $u_0$ for the same value
of $L$.  This is a problem since we expect a Wilson loop with a definite value
of $L$ to exhibit a unique physical behaviour \footnote{For large values of
$u_0$, the Wilson loop satisfies an area law as can easily be deduced from
(\ref{twoseventeen}) and (\ref{twoeighteen}).}.  If the boundary is placed at
large $\Lambda$, then a way out of this problem is to choose the branch
corresponding to the smaller value of $u_0$ as physical.  This is because this
branch is continuously connected to the limit in which the expected ordinary
AdS/CFT behaviour is obtained.

There is, however, another possibility, namely there may be a barrier,
$R\Lambda
\leq 1$.  In fact, we do find such a barrier for the purely space-like Wilson
loops in the next subsection.  If we do place the boundary at $R\Lambda \sim
1$,
then according to the formula (\ref{twoeighteena}) the minimum value of $L$ is
attained when $Ru_0 \sim R\Lambda \sim 1$.  We still have the Coulomb law
(\ref{twonineteen}) for $Ru_0<<1$.  However, when $u_0 \sim \Lambda$ the
approximation under which (\ref{twoseventeen}) and (\ref{twoeighteen}) were
calculated breaks down and so we must proceed differently.  This is what we do
in case(ii) below.  Before proceeding to that case, however, we mention that
the
metric in (\ref{twofourteen}) gives proper distances of order string scale when
coordinate distances in the non-commuting directions are of order the
non-commutativity scale if $R\Lambda \sim 1$.  For $R\Lambda>>1$ coordinate
distances much larger than the non-commutativity scale are needed to get proper
distances of order the string scale.  We believe that this observation provides
an important clue to having the possible barrier $R\Lambda \leq 1$.

\ni {\bf case(ii):$\displaystyle u_0 \mathop{<}_{\sim} \Lambda$}

As we have discussed above, this case becomes relevant only if there is a
barrier $R\Lambda \leq 1$.  A natural choice is to take $R\Lambda \sim 1$ since
at this point the metric in the non-commuting directions has a maximum and so
this $4$-dimensional slice can accommodate the maximum number of degrees of
freedom.  Also, this choice of ``boundary'' is consistent with the standard
AdS/CFT correspondence since $\Lambda$ goes to infinity in the commuting limit
$R \rightarrow 0$.  As mentioned above, with the ``boundary'' at $R\Lambda \sim
1$ we still get Coulomb law behaviour for $Ru_0<<1$.  However, for values of
$u_0$ close to $\Lambda=R^{-1}$ the behaviour is different.  In this case we
may
evaluate the integrals (\ref{twofifteen}) and (\ref{twosixteen}) in powers of
$(\Lambda/u_0-1)$.  We then get,
\bea
{L \over 2}={b+c \over bu_0} \sqrt{{\Lambda \over u_0}-1} \ + \cdots,
\label{twotwentyone}
\eea
and
\bea
S={T \sqrt\lambda \over \pi }(b+c)u_0 \sqrt{h_0} \sqrt{{\Lambda \over
u_0}-1} \
+ \cdots
\label{twotwentytwo}
\eea
where the dots refer to terms higher order in powers of $(\Lambda/ u_0-1)$.
Let
us, for the sake of definiteness, again take $v=0$.  Then, from
(\ref{twotwentyone}) and (\ref{twotwentytwo}) we get
\bea
S \approx {TL \over 2 \sqrt 2 \pi \theta},
\label{twotwentytwoa}
\eea
where we have used $Ru_0 \sim 1$ \footnote{Note, however, that $u_0$ cannot get
too close to $\Lambda$, otherwise $L$ will become too small for the present
analysis to be reliable.  An analysis of the expected sigma-model
corrections to
the leading semiclassical result shows that this restricts $L$ to be much
larger
than the non-commutativity scale.  This translates into the restriction
$(\Lambda/ u_0-1) >> 1/\sqrt{\lambda}$.  Since $\lambda$ is large this
condition
is easily satisfied.}.  We see that in this case there is a crossover to {\it
area law}.  The tension is controlled by the non-commutativity parameter.

In summary, then, the picture that has emerged, for the mixed type of loops
that
we have considered in this subsection, is that there is a crossover from
Coulomb
type of behaviour for large Wilson loops to area law for small Wilson loops,
provided the ``boundary'' is placed at $R\Lambda \sim 1$.  The tension involved
in the area law is determined by the non-commutativity scale, while the scale
involved in the crossover itself is quite likely the scale that enters the dual
supergravity solution.

\subsection{Rectangular Wilson loops with both directions non-commuting}

Rectangular Wilson loops with both directions non-commuting (and so
space-like) are obtained for negative values of $(1-v^2h)$. These are
possible only for $|v|>1$. Actually, for such values of $v$, $(v^2h-1)$
vanishes for $Ru=(v^2-1)^{1 \over 4}$. This value is large for large
values of $|v|$, and in the limit can be pushed to infinity. We will take
this limit in the following. Now, for positive values of $(v^2h-1)$, the
first term in the action (\ref{twoseven}) becomes imaginary and what we
have done effectively amounts to a Euclidean continuation, the Euclidean
action being
\bea
S_E={T \sqrt\lambda \over 2 \pi} \int d\sigma \bigg[
\sqrt{(v^2h-1)({u'}^2+u^4h)} -ivR^2u^4h \bigg].
\label{twotwentythree}
\eea
Taking the large $v$ limit and writing $vT \equiv L_2$, the action becomes
\bea
S_E={L_2 \sqrt\lambda \over 2 \pi} \int d\sigma \bigg[
\sqrt{h({u'}^2+u^4h)} -iR^2u^4h \bigg].
\label{twotwentyfour}
\eea
One could also have obtained the above action by starting from the
Euclidean version of the action in (\ref{twofour}) and choosing the string
configuration $x_2=\tau, \ x_3=\sigma, \ u=u(\sigma), \ x_0=x_1=0$. This
action also appears in the dual description of rectangular Wilson loops in
two mutually non-commuting directions in the fully non-commuting Euclidean
NCSYM theory \cite{MR,IIKK2}.

The Euclidean action (\ref{twotwentyfour}) is complex because of the
contribution of the B field. This gives rise to a phase in the path
integral which measures the magnetic flux passing through the string
world-sheet. In the following we will ignore this contribution to the
action and work only with the real part. Some a posteriori justification
for this will be given at the end.

At an extremum of the real part of the above action the quantity $$u^4h
{\sqrt h
\over \sqrt{{u'}^2+u^4h}}$$ is independent of $\sigma$.  This gives the
solution
\bea
u'={u^2 h^{3/2} \over u_0^2} \sqrt{u^4-u_0^4} \sqrt{1-R^8u_0^4u^4}.
\label{twotwentyfive}
\eea
We see that $u'=0$ at $u=u_0, 1/ R^2u_0$.  This reflects the symmetry of the
background metric under $u \rightarrow 1/R^2u$ in the non-commuting directions.
Because of this symmetry, it is sufficient for us to restrict ourselves to
$Ru_0
\leq 1$.  Then, we see from (\ref{twotwentyfive}) that starting from $u=u_0$,
the string turns back at $u=1/R^2u_0$.  It never goes all the way to infinitely
large $u$, though for very small values of $Ru_0$ it does go through a very
large region.

Let us now consider the contour of the Wilson loop at a $4$-dimensional
constant
$u$ slice at $u=\Lambda$.  As in the case of $u_0$, because of the symmetry $u
\rightarrow 1/R^2u$, it is sufficient to restrict to $R\Lambda \leq 1$.  In
fact, in the following we will take $R\Lambda \sim 1$.  Now, the length, $L_3$,
between the two ends of the string on this slice can be obtained from
(\ref{twotwentyfive}) and is given by
\bea
{L_3 \over 2}={1 \over u_0} \int_1^{\Lambda/u_0} dy {(1+R^4u_0^4y^4)^{3/2}
\over y^2 \sqrt{(y^4-1)(1-R^8u_0^8y^4)}}.
\label{twotwentysix}
\eea
The action evaluated on the solution is given by
\bea
S_E={L_2 \sqrt\lambda \over \pi}{u_0 \over h_0} \int_1^{\Lambda/u_0} dy
{y^2 \over \sqrt{(1+R^4u_0^4y^4)(y^4-1)(1-R^8u_0^8y^4)}}.
\label{twotwentyseven}
\eea
We now discuss the two cases that arise.

\ni {\bf case(i):$u_0 << \Lambda$}

It is clear from the expressions in (\ref{twotwentyfive}) and
(\ref{twotwentysix}) that in this case we can ignore the explicit dependence of
the integrands on $u_0$.  Thus, in this case one recovers the usual AdS/CFT
results, except for one difference.  In the case of ordinary AdS/CFT, the
perimeter term is large because it is proportional to a large cut-off.
Here, on
the other hand, the perimeter term is finite because the cut-off cannot be
larger than $R^{-1}$.  We will make important use of this observation in the
concluding section.

\ni {\bf case(ii):$\displaystyle u_0 \mathop{<}_{\sim} \Lambda$}

For $\displaystyle Ru_0 \mathop{<}_{\sim} 1$, we need to keep the explicit
dependence on $u_0$ in the integrands of (\ref{twotwentyfive}) and
(\ref{twotwentysix}). The integrals can be done by expanding in powers of
$(\Lambda/u_0-1)$. We get,
\bea
{L_3 \over 2} & \approx & \sqrt 2 R[{\pi \over 2}- {\rm
sin}^{-1}(1-{\epsilon \over \delta})+ \cdots],
\label{twotwentyeight} \\
S_E & \approx & {L_2 \sqrt\lambda \over \pi} {1 \over \sqrt 2 R}
[{\pi \over 2}-{\rm sin}^{-1}(1-{\epsilon \over \delta})+ \cdots],
\label{twotwentynine}
\eea
where $\epsilon \equiv {\Lambda/u_0 -1}, \ \delta \equiv {1-Ru_0}$ and the
dots refer to terms higher order in $\epsilon$ and $\delta$. Thus, we may
write
\bea
S_E \approx {L_2L_3 \over 4\pi \theta}.
\label{twothirty}
\eea
So in this case we find area law.  As before, the tension is determined purely
by the non-commutativity scale.  Note that the tension computed from
(\ref{twothirty}) is identical to that computed from (\ref{twotwentytwo}).
This
is because of the extra metric factor of $1/\sqrt 2$ in the proper length at
$R\Lambda=1$.  Note also that $L_3$ given by (\ref{twotwentynine}) is of order
$R$.  For smaller values of $L_3$ we need to take the other solution for which
$u'=0$ and the loop is spanned by a straight surface at $R\Lambda=1$.

To summarize, then, Wilson loops lying entirely in the non-commuting plane show
a crossover from Coulomb type of behaviour to area law as the loop size
decreases through the length scale $R$ which characterizes the supergravity
solution.

Finally, we will now provide some justification for neglecting the contribution
of the B field term in the action.  One can check from the solution
(\ref{twotwentyfive}) that the magnitude of this term relative to the action
(\ref{twotwentyseven}) is $R^2u_0^2h_0$.  So it is suppressed for $Ru_0<<1$.
For $\displaystyle {Ru_0} \mathop {<}_{\sim} 1$, $u$ is essentially constant
and
obviously the largest weight in the functional integral is for the world-sheet
with minimal area, which is essentially a straight surface on the slice at
$\Lambda \sim {1/R}$.  We expect that quantum fluctuations around this solution
are suppressed, but we have not checked that.  It is relevant to add here that
the B field term is actually absent for Wilson loops lying in mutually
commuting
directions in the fully non-commuting Euclidean NCSYM theory.

\section{\large Summary and concluding remarks}

We have performed a detailed analysis of Wilson loops in NCSYM theory in the
strong coupling region using the conjectured dual supergravity description.  We
have seen that the Wilson loops exhibit a crossover from Coulomb type of
behaviour for large Wilson loops, which essentially do not see any
non-commutativity, to area law for small Wilson loops, for which
non-commutativity is strong.  Perhaps the most interesting aspect of our
investigations is the likely existence of a length scale in NCSYM theory, where
the crossover from Coulomb behaviour to area law takes place.  The string
tension that emerges from the area law is determined by the non-commutativity
scale.  However, the crossover scale itself is probably much larger than this.
As we have seen, the crossover scale is likely to be the scale $R$ that enters
the dual supergravity description.

Our conclusions are based on the choice of the $4$-dimensional slice of the
dual
metric at $u \sim R^{-1}$ as the location of the ``boundary''.
The standard
choice of the location of the ``boundary'' at large $u$ gives Coulomb
behaviour for Wilson loops of arbitrary size.  In the case of commuting SYM
theory, Coulomb behaviour at all scales is guaranteed by conformal symmetry.
In the case of the NCSYM theory, there is no such symmetry.  Therefore, in this
case it is legitimate to find the area law, at least for those loops whose
size is
of order the non-commutativity scale, since one would expect
non-commutativity effects to show up for such loops.  Moreover, in the fully
non-commuting case (in which all the four coordinates are non-commuting), the
large $u$ region shrinks to a point, so it is hard to imagine how all the
degrees of freedom of the system can be located there.  Our choice of the
location of the ``boundary'' predicts interesting nonperturbative
behavior of NCYM, while being
compatible with the expectation that large loops should behave like in the
commutative SYM theory.

We stress here that our choice of putting the
``boundary'' at the length scale $R$ is not an ad-hoc choice.
In ordinary field theory, we can probe arbitrarily short distance
scale by colliding particles with very high momenta.
So any short distance cut-off must be removed in field theory.
In NCYM, the situation
is completely different. The counterpart to a local operator
with definite momentum is a Wilson line in NCYM.
In \cite{GHI}, the high energy limit of two point functions
of the Wilson lines is investigated.
Through their work, the average distance
between the Wilson lines can be estimated as
${(\lambda|\theta k|/|k|)}^{1/2}$ in
the weak coupling region.
It is the mimimum distance which can be probed
by physical observables in the theory.
In the uniformly large momentum region,
it is $O(\sqrt{\lambda\theta})$.
It may imply that the mimimum distance scale is indeed $R$
in the strong coupling region
if we adopt the standard replacement rule $\lambda
\rightarrow\sqrt{\lambda}$.
Therefore we argue that our proposal is consistent with
the existence of mimimum distance of $R$ in NCYM at strong coupling.

In fact
this investigation was partly motivated by the high energy behavior of the
multi-point correlators of Wilson lines.  The results we have obtained here are
consistent with it.
The comparison may be made between the normalized Wilson line
correlators in NCYM and our geometric prediction.
It may make sense to consider the normalized Wilson line operators
$W(C)/<W(C)W^{\dagger}(C)>^{1/2}$ since it is free from
the short distance divergences.
One can argue that in the extreme ultraviolet region,
for generic momenta, the
three-point and higher correlators should be related to large Wilson loops
lying
entirely in the non-commuting plane, with the length scale involved of order
$|k\theta|$.  Given this, we can make predictions for the extreme ultraviolet
behaviour of correlation functions, using the results we have obtained here.
The Wilson loop correlators
typically have a very large area since the length scale involved is
$|k\theta|$.
So these should exhibit Coulomb behaviour.  The Coulomb behaviour is
now actually dominated by the finite perimeter term we found in case(i) of
subsection 2.4.  So, this would predict a perimeter suppression for three-point
and higher correlators relative to the two-point function ,
which is qualitatively the result obtained in \cite{GHI}.

We therefore see the relevance of our proposal
to the high energy behavior of
the correlation functions of the Wilson lines in NCYM.
Finally, we mention that one can derive Schwinger-Dyson equations
\cite{FKKT,AZD,DW2} for NCSYM theories within the context of matrix models
\cite{BFSS,IKKT}. It would be interesting to obtain the behaviour we have
found here for Wilson loops as solutions to these equations.

\ni {\bf Acknowledgements}
\ni We would like to thank Satoshi Iso for discussions. One of us (AD) would
also like to thank Spenta Wadia for discussions.

\end{document}